\documentstyle[pra,aps]{revtex} 
\begin{document} 
  \draft 
  \title{Interactions from SSB
of scale symmetry: applications to problems of
\\
quintessence, galaxy dark matter 
and fermion family }
\author{E. I. Guendelman \thanks{guendel@bgumail.bgu.ac.il} and
A.  B.  Kaganovich \thanks{alexk@bgumail.bgu.ac.il}} 
\address{Physics Department, Ben Gurion University of the Negev, Beer Sheva 
84105, Israel}
\maketitle
\begin{abstract} 
We study a scale invariant two measures theory where a
dilaton field $\phi$ has no explicit potentials. The scale transformations
include a translation of a dilaton $\phi\rightarrow\phi +const$. The theory
 demonstrates
a new mechanism for generation of the exponential potential: in the conformal 
Einstein frame (CEF), after SSB of scale invariance, the  theory  develops
the 
exponential potential and, in general,
 non-linear  kinetic term is generated as well.
The scale symmetry does not allow the appearance of terms breaking the
exponential shape of the potential that solves the problem of the flatness
of the scalar field potential in the context of quintessential scenarios.
As examples, three different sets of the magnitudes of the parameters and
integration 
constants (SPaIC)
are presented where the theory permits to get interesting cosmological and 
astrophysical results in the analytic form. For one  SPaIC, the theory has
standard scaling
solutions for $\phi$ usually used in the context of the quintessential 
scenario. 
For the second SPaIC, the theory has  scaling solutions with 
equation of state $p_{\phi}=w\rho_{\phi}$ where 
$w$ is predicted to be restricted by $-1<w<-3/8$. For the third  SPaIC,
 the theory allows a static spherically symmetric 
particles in CEF that appears to suggest a new approach to the
family
problem of particle physics. It is automatically achieved that
for two of them, fermion masses are constants, gravitational
equations are
canonical and the "fifth force" is absent.
For the third  type of
particles, four fermionic interaction appears from SSB of scale
invariance.

   \end{abstract}

    \renewcommand{\baselinestretch}{1.6} PACS number(s): 98.80.Cq,
98.62.Gq, 11.30.Qc, 12.15.Pf

\pagebreak 

\section{Introduction} 
Recent observations imply that the Universe now is undergoing era of
acceleration\cite{P}. This is most naturally explained by the existence of
a vacuum energy which can be of the form of an explicit cosmological
constant. Alternatively, there may be a slow rolling scalar field, whose
potential (assumed to have zero asymptotic value) provides the negative
pressure required for accelerating the Universe. This is the
basic idea of the quintessence\cite{Wett1988NP668}. Some of the problems of the
quintessence scenario connected to the field theoretic grounds  of this
idea, are: 
i) what is the origin of the quintessence potential;
ii) why the asymptotic value of the potential vanishes (this is actually
the "old" cosmological constant problem\cite{CCP} )
iii) the needed flatness of the potential\cite{KLyth}.
iv) without the symmetry $\phi\rightarrow\phi +const$ it is very hard to
explain the absence of the long-range force if no fine tuning is
made\cite{Carroll,Dolgov}. But such a translation-like symmetry is usually
incompatible
with a nontrivial
potential.

One of the main aims of this paper is to show how the above problems
can be solved in the
context of the two measures
theories (TMT)\cite{GK1,GK2,GK3,GK4,G,G1,G2,K}. These kind of models are based on the
observation
that in a generally covariant formulation of the action principle one has
to integrate using an invariant volume element, which is not obliged to be
dependent of the metric. In GR, the volume element $\sqrt{-g}d^{4}x$ is
indeed generally coordinate invariant, but nothing forbids us from
considering the invariant volume element $\Phi d^{4}x$ where $\Phi$ is a
scalar density that could be independent of the metric\cite{GK1}.

If the measure $\Phi$ is allowed, we have seen in a number of
models\cite{GK2,GK3,GK4,G} that, in the conformal Einstein frame (CEF),
the equations of
motion have the canonical GR structure, but the scalar field potential
produced in  the CEF is such that zero vacuum energy for the
ground
state of the theory  is obtained without fine tuning, that is the "old"
cosmological constant problem can be solved\cite{GK4}.

If two measures are allowed, this opens new possibilities concerning scale
invariance\cite{G,G1,G2,K}. In this context
we study here a theory which is invariant under
scale transformations including also a translation-like symmetry for a
dilaton field of the form thought by Carroll\cite{Carroll}. For the case when the
original action does not contain dilaton potentials at all, it is found
that the
integration of the equation of motion corresponding to the measure $\Phi$
degrees of freedom, spontaneously breaks the scale symmetry and the
generation of a dilaton potential is a consequence of this 
spontaneous symmetry breaking (SSB). When
studying the theory in the CEF, it is demonstrated in
Sec. III that the
spontaneously induced dilaton potential has the form of the exponential
one and in addition, also non-linear kinetic terms appear. 

In Secs. IV and V, we discuss possible applications of the 
theory to cosmological and astrophysical questions when the dilaton field
is the dominant fraction of the matter: it is found that quintessential
solutions are possible and, for a different region of parameter space,
halo dark matter solutions are also possible.

In Sec. VI we show that
in the presence of fermions, the theory displays a
successful fermionic mass generation after the
SSB, and this is actually the second main aim of this paper. In the
regime when the
fermionic density is of
the order
typical for the normal particle physics (which in the laboratory
conditions is always much higher than the dilaton density ),
 there are constant fermion masses, gravitational equations
are
canonical and the "fifth force" is absent, - all this without any
additional restrictions on the parameters of the theory.
A possible explanation  to the "family puzzle" of particle physics
also appears naturally in the context of this model. For one of the 
families, a quartic fermion interaction appears as a result of the SSB of
scale symmetry.  

\section{Two Measures Theory (TMT)}

The main idea of these kind of theories\cite{GK1,GK2,GK3,GK4} is to reconsider the
basic structure of generally relativistic actions, which are usually taken
to be of the form
\begin{equation}
    S = \int d^{4}x\sqrt{-g}L
\label{SE}
\end{equation}
where $L$ is a scalar and $g=\det(g_{\mu\nu})$. The volume element
$d^{4}x\sqrt{-g}$ is an invariant entity. It is however possible to build
a different invariant volume element if another density, that is an object
having the same transformation properties  as $\sqrt{-g}$, is introduced.
For example, given four scalar fields  
$\varphi_{a}$, $a=1,2,3,4$ we can build the density
\begin{equation}
\Phi
=\varepsilon^{\mu\nu\alpha\beta}\varepsilon_{abcd}\partial_{\mu}\varphi_{a}
\partial_{\nu}\varphi_{b}\partial_{\alpha}\varphi_{c}\partial_{\beta}\varphi_{d}
\label{Phi}
\end{equation}
and then $\Phi d^{4}x$ is also an invariant object. Notice also that
$\Phi$ is a total derivative since
\begin{equation}
\Phi
=\partial_{\mu}(\varepsilon^{\mu\nu\alpha\beta}\varepsilon_{abcd}\varphi_{a}
\partial_{\nu}\varphi_{b}\partial_{\alpha}\varphi_{c}\partial_{\beta}
\varphi_{d})
\label{Phideriv}
\end{equation}
Therefore if we consider possible actions which use both $\Phi$ and
$\sqrt{-g}$ we are lead to TMT
\begin{equation}
    S = \int L_{1}\Phi d^{4}x +\int L_{2}\sqrt{-g}d^{4}x
\label{S}
\end{equation}

Since $\Phi$ is a total derivative, we see that a shift of $L_{1}$ by a
constant, $L_{1}\rightarrow L_{1}+const$, has the effect of adding to S
the integral of a total derivative , which does not change equations of
motion. Such a feature is not showed by the second piece of Eq. (\ref{S})
since $\sqrt{-g}$ is not a total derivative. It is clear then that the
introduction of a new volume element has consequences on the way we think
about the cosmological constant problem, since the vacuum energy is related
 to the coupling
of the volume element with the Lagrangian. How this relation is modified
when a new volume element is introduced, was discussed in
\cite{GK2,GK3,GK4}.

It has been shown that a wide class of TMT models\cite{GK4}, containing
among others a scalar field, can be formulated which are free of the "old"
cosmological constant problem. An  important  feature of those models
consists in the use of the
"first order formalism" where the
connection coefficients $\Gamma^{\lambda}_{\mu\nu}$, metric $g_{\mu\nu}$ and 
in our case also $\varphi_{a}$  and any
matter fields that may exist are treated as independent dynamical
variables. Any relations that they satisfy are a result of the
equations of motion.
The models allow the use of the so called conformal  Einstein frame (CEF)
where the equations of motion
have canonical GR form and the effective potential
has  an absolute minimum at zero value of the effective energy density 
without fine tuning. This was verified to be the case in all examples 
studied in Ref. \cite{GK4}, provided the action form (\ref{S}) is
preserved, 
where $L_{1}$ and $L_{2}$ are $\varphi_{a}$-independent. If this is so, 
an infinite symmetry appears\cite{GK4}: 
$\varphi_{a}\rightarrow\varphi_{a}+f_{a}(L_{1})$, where $f_{a}(L_{1})$
is an arbitrary function of  $L_{1}$.

\section{Scale invariant model with spontaneous symmetry breaking
\\
 giving rise to a potential}

If we believe that there are no fundamental scales in physics, we are lead 
to the notion of scale invariance. In the context of TMT, to
implement
global scale invariance one has to introduce a "dilaton" field\cite{G,G1}.
In this case the measure $\Phi$ degrees of freedom also can participate in 
the scale transformation\cite{G,G1}. In \cite{G,G1}, explicit potentials 
(of exponential form) which respect the symmetry were introduced.
Fundamental theories however, like string theories, etc. give most naturally
only massless particles, which means that only kinetic terms and no explicit 
potentials appear from the beginning naturally. Let us  therefore explore
a similar situation in the context of a scale invariant TMT model. We postulate
then the form of the action
\begin{equation}
S=\int d^{4}x\Phi e^{\alpha\phi/M_{p}}\left[
-\frac{1}{\kappa}R(\Gamma ,g)+\frac{1}{2}g^{\mu\nu}\phi_{,\mu} 
\phi_{,\nu}\right]
+\int d^{4}x\sqrt{-g}e^{\alpha\phi/M_{p}}\left[
-\frac{b_{g}}{\kappa}R(\Gamma ,g)+\frac{b_{k}}{2}g^{\mu\nu}\phi_{,\mu} 
\phi_{,\nu}\right]
\label{totac} 
\end{equation}
where we proceed in the first order formalism and 
$R(\Gamma,g)=g^{\mu\nu}R_{\mu\nu}(\Gamma)$,\hspace{0.25cm} 
$R_{\mu\nu}(\Gamma)=R^{\alpha}_{\mu\nu\alpha}(\Gamma)$ and 
$R^{\lambda}_{\mu\nu\sigma}(\Gamma)\equiv
\Gamma^{\lambda}_{\mu\nu,\sigma}+ 
\Gamma^{\lambda}_{\alpha\sigma}\Gamma^{\alpha}_{\mu\nu}- 
(\nu\leftrightarrow\sigma)$. 
By means of a redefinition of factors of $\phi$ and of $\Phi$
one can always normalize the kinetic term of $\phi$ and the $R$-term that 
go together with $\Phi$ as done in  (\ref{totac}). Once this is done, 
this freedom however is not present any more concerning the second part of 
the action going together with $\sqrt{-g}$. The appearance of the constants
$b_{g}$ and $b_{k}$ is a result of this. Concerning the possible
magnitudes of $b_{g}$ and $b_{k}$ we will here assume only that they are 
positive.

The action (\ref{totac}) is invariant under the scale transformations: 
\begin{equation}
    g_{\mu\nu}\rightarrow 
e^{\theta}g_{\mu\nu}, \quad
    \phi\rightarrow \phi-\frac{M_{p}}{\alpha}\theta ,\quad
\Gamma^{\sigma}_{\mu\nu}\rightarrow \Gamma^{\sigma}_{\mu\nu}, \quad
and \quad
\varphi_{a}\rightarrow \lambda_{a}\varphi_{a}\quad
where \quad \Pi\lambda_{a}=e^{2\theta}. 
\label{st} 
\end{equation}

Notice that (\ref{totac}) is the most general action of TMT 
invariant under the scale transformations (\ref{st}) where the Lagrangian
densities $L_{1}$ and $L_{2}$ are linear in the scalar curvature and 
quadratic in the space-time derivatives of the dilaton but {\it without
explicit potentials}. In Refs. \cite{G,G1}, actions of such type were 
discussed, but with explicit potentials and without kinetic term going
with $\sqrt{-g}$. A different definition
of the metric have been used also in \cite{G,G1} ($g^{\mu\nu}$ in
\cite{G,G1}
instead of
the combination
$e^{\alpha\phi/M_{p}}g^{\mu\nu}$ here) so that no factor 
$e^{\alpha\phi/M_{p}}$ appeared multiplying $\Phi$ in Ref.\cite{G,G1}.
Also it is possible to formulate a consistent scale invariant model
keeping only the simplest structure (namely, only the measure $\Phi$ is
used), provided $L_{1}$ contains 4-index field strengths and an
exponential potential for the dilaton\cite{G2}. Then SSB of the scale
invariance can lead to a quintessential potential\cite{G2}. Another type
of the field theory models with explicitly broken scale symmetry have been
studied in  Ref.\cite{K} where it is shown that the quintessential
inflation\cite{PV} type models can be obtained without fine tuning.

We examine now the equations of motion that arise from (\ref{totac}).
Varying the measure fields $\varphi_{a}$, we get
\begin{equation} 
A^{\mu}_{a}\partial_{\mu}[-\frac{1}{\kappa}R(\Gamma ,g)+
\frac{1}{2}g^{\alpha\beta}\phi_{,\alpha}\phi_{,\beta}]=0
\label{7}
\end{equation}
\begin{equation} 
A^{\mu}_{a}=\varepsilon^{\mu\nu\alpha\beta}\varepsilon_{abcd}
\partial_{\nu}\varphi_{b}\partial_{\alpha}\varphi_{c}
\partial_{\beta}\varphi_{d}.
\label{8}
\end{equation}

Since
$Det (A^{\mu}_{a})
= \frac{4^{-4}}{4!}\Phi^{3}$ it follows that if $\Phi\neq 0$, 
\begin{equation}
 -\frac{1}{\kappa}R(\Gamma,g)+
\frac{1}{2}g^{\mu\nu}\phi_{,\mu}\phi_{,\nu}=sM^{4}=const,
\label{varphi}
\end{equation}
where $s=\pm 1$ and $M$ is a constant of the dimension of mass.
It can be noticed that the appearance of a nonzero integration 
constant $sM^4$ spontaneously breaks the scale invariance (\ref{st}).

The variation of $S$ with respect to $g^{\mu\nu}$ yields
\begin{equation}
-\frac{1}{\kappa}R_{\mu\nu}(\Gamma)(\Phi +b_{g}\sqrt{-g})+
\frac{1}{2}\phi_{,\mu}\phi_{,\nu}
(\Phi +b_{k}\sqrt{-g})-
\frac{1}{2}\sqrt{-g}g_{\mu\nu}
\left[-\frac{b_{g}}{\kappa}R(\Gamma ,g)+
\frac{b_{k}}{2}g^{\alpha\beta}\phi_{,\alpha}\phi_{,\beta}\right]=0
\label{varg}
\end{equation}

Contracting Eq. (\ref{varg}) with $g^{\mu\nu}$, solving for $R(\Gamma ,g)$
and inserting into Eq. (\ref{varphi}) we obtain the constraint
\begin{equation}
M^{4}(\zeta -b_{g})e^{-\alpha\phi/M_{p}}+
\frac{\Delta}{2}g^{\alpha\beta}\phi_{,\alpha}\phi_{,\beta}
=0,
\label{con1}
\end{equation}
where the scalar $\zeta$ is defined as
\begin{equation}
\zeta \equiv\frac{\Phi}{\sqrt{-g}}
\label{zeta}
\end{equation}
and $\Delta =b_{g}-b_{k}$.

Varying the action with respect to $\phi$ and using Eq. (\ref{varphi})
we get
\begin{equation} 
(-g)^{-1/2}\partial_{\mu}\left[(\zeta +b_{k})e^{\alpha\phi/M_{p}}
\sqrt{-g}g^{\mu\nu}\partial_{\nu}\phi)\right]-
\frac{\alpha}{M_{p}}\left[M^{4}(\zeta +b_{g})
-\frac{\Delta}{2}g^{\alpha\beta}\phi_{,\alpha}\phi_{,\beta}
e^{\alpha\phi/M_{p}}\right]=0
\label{se}
\end{equation}

Considering the term containing connection $\Gamma^{\lambda}_{\mu\nu}$, that is 
$R(\Gamma ,g)$, we see that it can be written as
\begin{equation}
S_{\Gamma}=-\frac{1}{\kappa}\int \sqrt{-g}e^{\alpha\phi/M_{p}}(\zeta +b_{g})
g^{\mu\nu}R_{\mu\nu}(\Gamma) =
-\frac{1}{\kappa}\int \sqrt{-\tilde{g}}\tilde{g}^{\mu\nu}R_{\mu\nu}(\Gamma),
\label{Gamac}
\end{equation}
where $\tilde{g}_{\mu\nu}$ is determined by the conformal transformation
\begin{equation}
\tilde{g}_{\mu\nu}=e^{\alpha\phi/M_{p}}(\zeta +b_{g})g_{\mu\nu}
\label{ct}
\end{equation}

It is clear then that the variation of $S_{\Gamma}$ with respect to $\Gamma$
will give the same result expressed in terms of $\tilde{g}_{\mu\nu}$
as in the similar GR problem in Palatini formulation. Therefore, if 
$\Gamma^{\lambda}_{\mu\nu}$ is taken to be symmetric in $\mu , \nu$, then
in terms of the metric $\tilde{g}_{\mu\nu}$, the connection coefficients
$\Gamma^{\lambda}_{\mu\nu}$ are Christoffel's connection coefficients of 
the Riemannian space-time with the metric
$\tilde{g}_{\mu\nu}$:
\begin{equation}
\Gamma^{\lambda}_{\mu\nu}=
\{
^{\lambda}_{\mu\nu}\}|_{\tilde{g}_{\mu\nu}}=\frac{1}{2}\tilde{g}^{\lambda\alpha}
(\partial_{\nu}\tilde{g}_{\alpha\mu}+\partial_{\mu}\tilde{g}_{\alpha\nu}-
\partial_{\alpha}
\tilde{g}_{\mu\nu}).
\label{Gama}
\end{equation}

So, it appears that working with $\tilde{g}_{\mu\nu}$, we recover a Riemannian 
structure  for space-time. We will refer to this as the conformal Einstein
frame (CEF). Notice that $\tilde{g}_{\mu\nu}$ is invariant under the scale 
transformations (\ref{st}) and therefore the spontaneous breaking of the global
scale symmetry (see Eq. (\ref{varphi}) and discussion after it) is reduced, in CEF, 
to the spontaneous breaking of the shift symmetry $\phi\rightarrow\phi +const$
for the dilaton field.
In this context, it is interesting to notice 
that Carroll\cite{{Carroll}} pointed
to the possible role of the shift symmetry  for a scalar field in the
resolution of the long range force problem of the quintessential scenario.

Equations (\ref{varg}) and (\ref{se}) in CEF
take the following form:
\begin{equation}
G_{\mu\nu}(\tilde{g}_{\alpha\beta})=\frac{\kappa}{2}T_{\mu\nu}^{eff}
 \label{gef}
\end{equation}
\begin{equation}
T_{\mu\nu}^{eff}=\frac{1}{2}\left(1+\frac{b_{k}}{b_{g}}\right)
\left(\phi_{,\mu}\phi_{,\nu}-K\tilde{g}_{\mu\nu}\right)-
\frac{\Delta^{2}Ke^{2\alpha\phi/M_{p}}}{2b_{g}M^{4}}
\left(\phi_{,\mu}\phi_{,\nu}-\frac{1}{2}K\tilde{g}_{\mu\nu}\right)+
\tilde{g}_{\mu\nu}\frac{sM^{4}}{4b_{g}}e^{-2\alpha\phi/M_{p}}
 \label{Tmn}
\end{equation}
\begin{eqnarray}
\left(b_{g}+b_{k}-\frac{\Delta^{2}}{M^{4}}Ke^{2\alpha\phi/M_{p}}\right)
\left[(-\tilde{g})^{-1/2}\partial_{\mu}
(\sqrt{-\tilde{g}}\tilde{g}^{\mu\nu}\partial_{\nu}\phi ) +
\tilde{g}^{\alpha\beta}\partial_{\alpha}\phi\partial_{\beta}
\ln|\frac{1}{2}\left(1+\frac{b_{k}}{b_{g}}\right)-
\frac{\Delta^{2}}{2b_{g}M^{4}}Ke^{2\alpha\phi/M_{p}}|\right]
 \nonumber\\
+\frac{\alpha\Delta^{2}}{M_{p}M^{4}}K^{2}e^{2\alpha\phi/M_{p}}
-\frac{\alpha M^{4}}{M_{p}}e^{-2\alpha\phi/M_{p}}=0
\label{phief}
\end{eqnarray}

Here $K\equiv\frac{1}{2}\tilde{g}^{\alpha\beta}\phi_{,\alpha}
\phi_{,\beta}$,
\quad $G_{\mu\nu}(\tilde{g}_{\alpha\beta})$
is the Einstein tensor in the Riemannian space-time with metric
$\tilde{g}_{\mu\nu}$ and  the constraint (\ref{con1}) have been used which 
in  CEF
takes the form
\begin{equation}
\zeta =b_{g}\frac{M^{4}-\Delta Ke^{2\alpha\phi/M_{p}}}
{M^{4}+\Delta Ke^{2\alpha\phi/M_{p}}}
\label{con2}
\end{equation}

Notice that in $T_{\mu\nu}^{eff}$ we can recognize an effective potential 
\begin{equation}
V_{eff}=\frac{sM^{4}}{4b_{g}}e^{-2\alpha\phi/M_{p}}
\label{Veff}
\end{equation}
 which appears in spite
of the fact that no explicit potential term was introduced in the original 
action
(\ref{totac}). As we see, the existence of $V_{eff}$ is associated with 
the constant $sM^{4}$, appearance of which spontaneously  breaks the scale 
invariance. This is actually a new mechanism for generating the
exponential potential\footnote{See for comparison Refs.
\cite{exppot1,exppot2,exppot3}
and a general discussion in Ref. \cite{FJ}}. 

Notice also that if $b_{g}\neq b_{k}$, the effective energy-momentum 
$T_{\mu\nu}^{eff}$ as well as the dilaton equation of motion contain
 the non-canonical terms nonlinear\footnote{Another  origins for
non-linear kinetic terms, known in the literature\cite{WP},
 are higher order gravitational corrections in
string and supergravity theories} in gradients of the dilaton $\phi$. 
It will be very important that the 
non-canonical in $\phi_{,\alpha}$ terms are multiplied by a very 
specific exponential of $\phi$. As we will see, these non-canonical terms 
may be responsible for the most interesting scaling solutions.
In the context of FRW cosmology, this structure
provides conditions for quintessential solutions if $s=1$. In the case of static
solutions, it garanties the existence of solutions which could play the role
 of the halo dark matter of galaxies provided that $s=-1$.

\section{Scaling solutions}

In the context of a spatially flat FRW cosmology with a metric $ds^{2}_{eff}=
\tilde{g}_{\mu\nu}dx^{\mu}dx^{\nu} =dt^{2}-a^{2}(t)(dx^{2}+dy^{2}+dz^{2})$,
the equations (\ref{gef})-(\ref{phief}), with the choice $s=+1$, become:
\begin{equation}
H^{2}=\frac{1}{3M_{p}^{2}}\rho_{eff}(\phi)
\label{FRWgr}
\end{equation}
\begin{eqnarray}
\left(b_{g}+b_{k}-
\frac{\Delta^{2}}{2M^{4}}\dot{\phi}^{2}e^{2\alpha\phi/M_{p}}\right)
\left[\ddot{\phi}+3H\dot{\phi}+
\dot{\phi}\partial_{t}\ln|\frac{1}{2}\left(1+\frac{b_{k}}{b_{g}}\right)-
\frac{\Delta^{2}}{4b_{g}M^{4}}\dot{\phi}^{2}e^{2\alpha\phi/M_{p}}|\right]
\\
+\frac{\alpha\Delta^{2}}{4M^{4}M_{p}}\dot{\phi}^{4}e^{2\alpha\phi/M_{p}}
-\frac{\alpha M^{4}}{M^{p}}e^{-2\alpha\phi/M_{p}}=0
\label{FRWphi}
\end{eqnarray}
where the energy density of the dilaton field is
\begin{equation}
\rho_{eff}(\phi)=\frac{1}{4}\left(1+\frac{b_{k}}{b_{g}}\right)\dot{\phi}^{2}
-\frac{3\Delta^{2}}{16b_{g}M^{4}}\dot{\phi}^{4}e^{2\alpha\phi/M_{p}}+ 
\frac{M^{4}}{4b_{g}}e^{-2\alpha\phi/M_{p}}
\label{rho}
\end{equation}
and the pressure
\begin{equation}
p_{eff}(\phi)=\frac{1}{4}\left(1+\frac{b_{k}}{b_{g}}\right)\dot{\phi}^{2}
-\frac{\Delta^{2}}{16b_{g}M^{4}}\dot{\phi}^{4}e^{2\alpha\phi/M_{p}}-
 \frac{M^{4}}{4b_{g}}e^{-2\alpha\phi/M_{p}}
\label{p}
\end{equation}

One can see that Eqs. (\ref{FRWgr})-(\ref{rho}) allow  solutions
of a familiar quintessential form\cite{Wett1988NP668,FJ}
\begin{equation}
\phi(t)=\frac{M_{p}}{2\alpha}\phi_{0}+\frac{M_{p}}{\alpha}\ln (M_{p}t)
\label{phiq}
\end{equation}
\begin{equation}
a(t)=t^{\gamma}
\label{aq}
\end{equation}
which provides scaling behaviors of the dilaton energy density
\begin{equation}
\rho_{\phi}\propto 1/a^{n}.
\label{scalrho}
\end{equation}
The important role for possibility of such solutions belongs to
the remarkable feature of the nonlinear terms in 
Eqs. (\ref{FRWgr})-(\ref{rho})  that appear only in the combination
$\dot{\phi}^{2}e^{2\alpha\phi/M_{p}}$ which remains constant for
the solutions (\ref{phiq}) and (\ref{aq}): 
\begin{equation}
\dot{\phi}^{2}e^{2\alpha\phi/M_{p}}=const
\label{constnonlinear}
\end{equation}

Eqs. (\ref{phiq}) and (\ref{aq}) describe solutions of 
Eqs. (\ref{FRWgr})-(\ref{rho}) if
\begin{equation}
\gamma =\frac{b_{g}+b_{k}-y}{4b_{g}\alpha^{2}}
\label{gam}
\end{equation}
where 
\begin{equation}
y\equiv\frac{\Delta^{2}M_{p}^{4}e^{\phi_{0}}}{2M^{4}\alpha^{2}} 
\label{y}  
\end{equation}
is a solution of the cubic equation
\begin{equation}
y^{3}-2(b_{g}+b_{k}-b_{g}\alpha^{2})y^{2}+(b_{g}+b_{k})(b_{g}+b_{k}-
\frac{4}{3}b_{g}\alpha^{2})y-\frac{2}{3}b_{g}\alpha^{2}\Delta^{2}=0. 
\label{eqy}  
\end{equation}

Up to now we did not make any assumptions about parameters of the
theory. We will suppose that $b_{g}$ and $b_{k}$ are positive.
 One can notice immediately that if $b_{k}=b_{g}$ then  
Eqs. (\ref{FRWgr})-(\ref{p}) describe the FRW cosmological model
 in the context of the standard GR when the minimally coupled
scalar field $\phi$ with the
potential $\frac{M^{4}}{4b_{g}}e^{-2\alpha\phi/M_{p}}$ is the only
source of gravity. 

Another interesting possibility consists of the
assumption that
\begin{equation}
b_{k}\ll b_{g}
\label{bkey}  
\end{equation}
Then ignoring corrections of the order of $b_{k}/b_{g}$,
the solutions  of Eq. (\ref{eqy}) are
\begin{equation}
y_{1}=b_{g}
\label{soly1}  
\end{equation}
 \begin{equation}
y_{2}=\frac{b_{g}}{2}\left[1-2\alpha^{2}+\sqrt{4\alpha^{4}-
\frac{20}{3}\alpha^{2}+1}\right]
\label{soly2}  
\end{equation}
\begin{equation}
y_{3}=\frac{b_{g}}{2}\left[1-2\alpha^{2}-\sqrt{4\alpha^{4}-
\frac{20}{3}\alpha^{2}+1}\right]
\label{soly3}  
\end{equation}

The solution $y_{1}$ corresponds to the static universe 
($\gamma =0$ and $a(t)=const$) supported by the slow rolling scalar
field $\phi$, Eq. (\ref{phiq}). However, taking into account
corrections of the order $b_{k}/b_{g}$ to $y_{1}$ we will get
$\gamma\propto {\cal O}(b_{k}/b_{g})$.

Solutions $y_{2}$ and $y_{3}$ exist and are positive (see the 
definition (\ref{y})) only if
\begin{equation}
\alpha^{2}\leq \frac{1}{6}
\label{alpha}
\end{equation}

The solution $y_2$ corresponds to the values of the parameter
$\gamma$ monotonically varying from $\gamma_{min}=2/3$
up to $\gamma =1$ as $\alpha^2$ changes from 0 up to $1/6$.

The most interesting solution is given by $y_{3}$ that provides
the values of the parameter
$\gamma$ monotonically varying from $\gamma_{min}= 1$
up to $\infty$ as $\alpha^2$ changes from $1/6$ up to zero.
In this case, Eqs. (\ref{phiq})-(\ref{aq}) describe an
accelerated  universe for all permissible values of $\alpha^2$ 
and the energy density of the dilaton field scales as in Eq. 
(\ref{scalrho}) with monotonically varying $n$, $2\geq n\geq 0$ as
 $\alpha^2$ changes from $1/6$ up to zero. For the dilatonic matter
equation-of-state $p=w\rho$ we get 
\begin{equation}
-1\leq w\leq -32/39\approx -0.82
\label{w}
\end{equation}

In the conclusion of this section let us revert to one of the problems of
the quintessence discussed in Introduction, namely to the flatness
problem\cite{KLyth}. This is a question of the field theoretic
basis
for the choice of the flat enough potential. In fact, Kolda and Lyth noted
\cite{KLyth} that an extreme fine tuning is needed in order to prevent the
contribution from another possible terms breaking the flatness of the
potential (see also for a review by Binetruy in Ref.\cite{CCP}). In the
theory we
study here,
there is a symmetry (scale
symmetry (\ref{st})) which forbids the appearance of such dangerous
contributions into $V_{eff}$, at least on the classical level.  One can
hope that the soft breaking of the scale symmetry garanties that the
symmetry breaking quantum corrections to the classical effective 
potential (\ref{Veff}) will be small.

\section{Possibility for halo-dark-matter-like solutions 
\\
from 
spontaneous breaking of scale symmetry}

The idea that scalar field(s) configuration can give a "halo dark
matter" has been explored in the literature. For example, using 
a variation of the Barriola and Vilenkin topologically nontrivial
global monopol\cite{BV}, which provides with an energy density behaving
as $1/r^{2}$, Nucamendy, Selgado and Sudarsky\cite{Nuc} were able to 
find a solution of the halo dark matter  problem.

Another interesting and more simple model for dilatonic halo dark matter
have been studied by Matos, Guzman and Nunez\cite{Guz}. They
showed that a {\it single} spherically symmetric  scalar field with
exponential potential of the form (\ref{Veff}) could serve as a dark
matter in galaxies provided that the overall sign of the potential
(\ref{Veff}) is opposite to that used in quintessence cosmology
\cite{Wett1988NP668,FJ}. The physical origin of this opposite sign is a
serious problem of the model \cite{Guz}.

We will now see that the scale invariant model with a single scalar
field discussed in this paper can,  after SSB of scale symmetry,
give rise to the halo dark matter type solutions similar to those
studied in Ref.
\cite{Guz}. The appearance of the overall negative sign in the
potential is here a result of the choice of the negative integration
constant in Eq.(\ref{varphi}), i.e. $s=-1$ in Eq.(\ref{Veff}).

Let us consider Eqs. (\ref{gef})-(\ref{phief}) for the static spherically
case  
\begin{equation}
ds^{2}=B(r)dt^{2}-A^{-1}dr^{2}-r^{2}(d\theta^{2}+
sin^{2}\theta d\varphi^{2})
\label{interhal}
\end{equation}

Motivated from the cosmological solution where condition
(\ref{constnonlinear}) was satisfied, we now look for solutions when
\begin{equation}
\phi^{\prime}e^{\alpha\phi/M_{p}}=const
\label{consthalononlin}
\end{equation}
where $\phi^{\prime}\equiv\frac{d\phi}{dr}$. Then similar to the solutions
of Ref.\cite{Guz}, we get
\begin{equation}
\phi =\frac{M_{p}}{\alpha}\ln(r/r_{0})+\frac{M_{p}}{2\alpha}\phi_{1},
\quad
B(r)=r^{2l}, \quad A=const
\label{halosol}
\end{equation}
with the following equations for parameters $l$ and $\frac{xA}{r_{0}^{2}}$
where $x=\frac{M_{p}^{2}e^{\phi_{1}}}{\alpha^{2}M^{4}}$ \ :
\begin{equation}
l =\frac{\pi}{b_{g}\alpha^{2}}\left[2(b_{g}+b_{k})+
s\Delta^{2}\frac{xA}{r_{0}^{2}}\right]
\label{lambda}
\end{equation}
 \begin{equation}
(1+2l)\Delta^{2}\left(\frac{xA}{r_{0}^{2}}\right)^{2}+
4s(1+l)(b_{g}+b_{k})\frac{xA}{r_{0}^{2}}+4=0
\label{xAC2}
\end{equation}
In contrast to the cosmological applications of the theory studied in
the previous section, a positive solution for $\frac{xA}{r_{0}^{2}}$ 
of Eq. (\ref{xAC2})
exists only when $s=-1$.

Solution (\ref{halosol}) can be used for description of the halo
dark matter if $l\ll 1$ (see Ref. \cite{Nuc,Guz}). Having this in
mind 
and taking into account that $|\Delta|/(b_{g}+b_{k})<1$, we get
from Eqs. (\ref{lambda}) and (\ref{xAC2})
\begin{equation}
l\approx\frac{\pi}{\alpha^{2}}(1+\frac{b_{k}}{b_{g}})\left[2-
\left(\frac{b_{g}-b_{k}}{b_{g}+b_{k}}\right)^{2}\right]. 
\label{lambdavalue}
\end{equation}
Notice that for the particular case when $b_{g}=b_{k}$, the solution 
(\ref{lambdavalue})
gives $l=8\pi/\alpha ^{2}$ which coincides with the appropriate relation
between $l$ and exponent in the potential of Ref. \cite{Guz}.
 
This means that the halo dark matter type solution can be achieved
if $\alpha^{2}$ is large enough
\begin{equation}
\alpha^{2}\gg 4\pi
\label{alpharestrict}
\end{equation} 
in contrast to the condition (\ref{alpha}) needed for the existence of
the cosmological scaling solutions.

\section{A note on quantization}

If $\Delta\neq 0$ then one can see from Eq. (\ref{rho}) that there is  
a possibility of negative energy contribution from the space-time 
derivatives of the dilaton. This raises of course the suspicion that the
quantum theory may contain ghosts. 
Let us check this question when considering small perturbations around
the backgrounds studied in Secs. IV and V.

To see this, let us calculate the canonically conjugate momenta to $\phi$,
starting from the original action (\ref{totac}) and expressing it in
terms of 
the variables defined in CEF, Eq. (\ref{ct}):
\begin{equation}
\pi_{\phi}=\frac{1}{2b_{g}}\left(b_{g}+b_{k}-
\frac{\Delta^{2}}{sM^{4}}Ke^{2\alpha\phi/M_{p}}\right)\sqrt{-\tilde{g}}
\tilde{g}^{00}\dot{\phi}
\label{canconj}
\end{equation} 

As we have seen in Secs. IV and V, both the cosmological scaling solutions
and the halo-like solutions provide backgrounds where 
$Ke^{2\alpha\phi/M_{p}}=const$. Moreover, it is easy to see that 
for the scaling solutions
\begin{equation}
\pi_{\phi}=\frac{1}{2b_{g}}\left(b_{g}+b_{k}-y\right)a^{3}\dot{\phi}=
2\alpha^{2}\gamma a^{3}\dot{\phi},
\label{canconjscaling}
\end{equation} 
where $\gamma$ and $y$ are defined by Eqs. (\ref{gam}) and (\ref{y}).
We have seen that for scaling solutions studied in Se. IV, 
 $\gamma$ gets positive values. Therefore we conclude that in such
backgrounds $\pi_{\phi}$ and $\dot{\phi}$ have the same sign, that 
guaranties a ghost-free quantization.
 The only exclusion is the particular case when $b_{k}=0$, $y=b_{g}$. As
we
have seen, such solution describes a static universe. In this case
the canonically conjugate momenta $\pi_{\phi}=0$ and therefore it appears
that in this vacuum there are no particles associated with the scalar
field $\phi$.

For the   background constituted by the halo dark matter solution, 
Sec. V, we obtain
\begin{equation}
\pi_{\phi}=\frac{1}{8b_{g}}\left[4(b_{g}+b_{k})-
\Delta^{2}\frac{xA}{r_{0}^{2}}\right]A^{-1}r^{2-l}|\sin\theta |\dot{\phi},
\label{canconjhalo}
\end{equation} 
 where $x$ is defined in the text before Eq. (\ref{lambda}).
The coincidence of the signs of  $\pi_{\phi}$ and $\dot{\phi}$ 
follows then from Eq. (\ref{lambda}) as $s=-1$ and
positivity of $l$.

Thus, at least for the the physically interesting cases studied in Secs.
IV and V, the problem of ghosts does not appear.
 
\section{Inclusion of fermionic matter consistent with scale invariance
\\
and the "family birth effect"}

In general scalar-tensor theories, particle masses depend on time, when
the theory is studied in the frame where Newton's constant is really a
constant. However, for all the fermionic matter observed in the universe,
the cosmological variation of particle masses (including those of electrons)
is highly constrained. We want to show now how the theory presented in 
this paper avoids this problem and also the so called fifth force problem,
in spite of the need to include exponential couplings of the dilaton field
 to fermionic matter  in order to ensure global scale invariance.

To describe fermions, normally one uses the vierbein ($e_{a}^{\mu}$) and 
spin-connection ($\omega_{\mu}^{ab}$) formalism where the metric is given by 
$g^{\mu\nu}=e^{\mu}_{a}e^{\nu}_{b}\eta^{ab}$ and the scalar curvature is
$R(\omega ,e) =e^{a\mu}e^{b\nu}R_{\mu\nu ab}(\omega)$ where
\begin{equation} 
R_{\mu\nu ab}(\omega)=\partial_{\mu}\omega_{\nu ab}
+\omega_{\mu a}^{c}\omega_{\nu cb}
-(\mu\leftrightarrow\nu).
        \label{B}
\end{equation}

Following the general idea of the model, we now treat the geometrical 
objects $e_{a}^{\mu}$, $\omega_{\mu}^{ab}$, the measure fields $\varphi_{a}$,
as well as the dilaton $\phi$ and the fermionic fields as independent variables.
In this formalism, the natural generalization of the action (\ref{totac})
keeping the general structure (\ref{S}), 
when a fermion field $\Psi$ is also present and which also respect scale 
invariance is the following:
\begin{equation}
L_{1}=e^{\alpha\phi /M_{p}}{-\frac{1}{\kappa}R(\omega ,e)+
\frac{1}{2}g^{\mu\nu}\phi,_{\mu}\phi,_{\nu}
+\frac{i}{2}\overline{\Psi}\left[\gamma^{a}e_{a}^{\mu}
(\overrightarrow{\partial}_{\mu}+\frac{1}{2}\omega_{\mu}^{cd}\sigma_{cd})
-(\overleftarrow{\partial}_{\mu}-\frac{1}{2}\omega_{\mu}^{cd}\sigma_{cd})
\gamma^{a}e_{a}^{\mu}\right]\Psi 
-m\overline{\Psi}\Psi e^{\frac{1}{2}\alpha\phi/M_{p}}}
 \label{L1gr+mat}
 \end{equation}
\begin{equation}
L_{2}=e^{\alpha\phi /M_{p}}\left[-\frac{b_{g}}{\kappa}R(\omega ,e)+
\frac{b_{k}}{2}g^{\mu\nu}\phi,_{\mu}\phi,_{\nu}
-hm\overline{\Psi}\Psi e^{\frac{1}{2}\alpha\phi/M_{p}}\right]
 \label{L2gr+mat}
 \end{equation}  
The action (\ref{S}) with such $L_{1}$ and $L_{2}$ is invariant under 
the scale transformations
\begin{eqnarray}
    e_{\mu}^{a}\rightarrow e^{\theta /2}e_{\mu}^{a}, \quad
\omega^{\mu}_{ab}\rightarrow \omega^{\mu}_{ab}, \quad
\varphi_{a}\rightarrow \lambda_{a}\varphi_{a}\quad
where \quad \Pi\lambda_{a}=e^{2\theta} 
\nonumber
\\ 
\phi\rightarrow \phi-\frac{M_{p}}{\alpha}\theta ,\quad
\Psi\rightarrow e^{-\theta /4}\Psi, \quad 
\overline{\Psi}\rightarrow  e^{-\theta /4} \overline{\Psi}. 
\label{stferm} 
\end{eqnarray}

Notice that two types of fermionic "mass-like terms" which 
respect scale invariance have been introduced. In contrast, for
simplisity, we have restricted the coupling of the fermionic kinetic 
term to the measure $\Phi$ only.

We can immediately obtain the equations of motion. From these going
through
similar steps to those performed in Sec. III, a constraint follows again
which replaces (\ref{con1}) and which contains now a contribution from the
fermions. The spin-connection can be found by the variation of
$\omega^{\mu}_{ab}$. 

Similar to what we learned from the treatment of Sec.III, we can consider
the theory in the CEF which in this case involves
also a transformation of the fermionic fields:
\begin{eqnarray}
\tilde{g}_{\mu\nu}=e^{\alpha\phi/M_{p}}(\zeta +b_{g})g_{\mu\nu}, \quad  
\tilde{e}_{a\mu}=e^{\frac{1}{2}\alpha\phi/M_{p}}(\zeta
+b_{g})^{1/2}e_{a\mu},
\nonumber
\\
\Psi^{\prime}=e^{\frac{1}{4}\alpha\phi/M_{p}}(\zeta +b_{g})^{-1/4}\Psi .
\label{ctferm}
\end{eqnarray}

Notice that variables $\tilde{g}_{\mu\nu}$, $\tilde{e}_{a\mu}$,
$\Psi^{\prime}$ and $\overline{\Psi}^{\prime}$ are in fact invariant under
the scale transformations (\ref{stferm}). In the CEF 
the only field which still has a non trivial transformation property is
the dilaton $\phi$ which gets shifted (according to (\ref{stferm})).
Thus, the presence of fermions does not change a conclusion made in Sec.III
after Eq.(\ref{Gama}): the spontaneous breaking of the 
scale symmetry  is reduced, in the CEF, 
to the spontaneous breaking of the shift symmetry $\phi\rightarrow\phi +const$
for the dilaton field.

In terms of  $\tilde{e}_{a\mu}$,
$\Psi^{\prime}$, $\overline{\Psi}^{\prime}$ and $\phi$, the constraint 
which now replaces (\ref{con2}) and which contains now a contribution from
 the fermions is
\begin{equation}
(\zeta -b_{g})M^{4}e^{-2\alpha\phi/M_{p}}+
\Delta (\zeta +b_{g})K+
F(\zeta)(\zeta +b_{g})^{1/2}
m\overline{\Psi}^{\prime}\Psi^{\prime}=0.
\label{confermEin}
\end{equation}
where we have chosen $s=+1$ for definiteness and the function $F(\zeta)$ is
defined by
\begin{equation}
F(\zeta)\equiv \frac{1}{2}\left(\zeta +\frac{2b_{g}h}{\zeta}+3h\right)
\label{F}
\end{equation}

The dilaton and the fermion field equations are respectively
\begin{eqnarray}
(\zeta +b_{k})
\left[(-\tilde{g})^{-1/2}\partial_{\mu}
(\sqrt{-\tilde{g}}\tilde{g}^{\mu\nu}\partial_{\nu}\phi ) +
\tilde{g}^{\alpha\beta}\partial_{\alpha}\phi\partial_{\beta}
\ln {\ |}\frac{\zeta +b_{k}}{\zeta +b_{g}}{\ |}\right]  
 \nonumber\\
+\frac{\alpha\Delta}{M_{p}}K
-\frac{\alpha M^{4}}{M_{p}}e^{-2\alpha\phi/M_{p}}+
\frac{\alpha m}{M_{p}\sqrt{\zeta +b_{g}}}F(\zeta)
\overline{\Psi}^{\prime}\Psi^{\prime}=0.
\label{phief+ferm}   
\end{eqnarray}
\begin{equation}
\left\{i\left [\tilde{e}_{a}^{\mu}\gamma^{a}\left
(\partial_{\mu}-ieA_{\mu}\right )+
\gamma^{a}\tilde{C}^{b}_{ab}
+\frac{i}{4}\tilde{\omega}_{\mu}^{cd}\varepsilon_{abcd}\gamma^{5}\gamma^{b}
\tilde{e}^{a\mu}\right ]
-\frac{m}{\sqrt{\zeta +b_{g}}}\left(1+\frac{h}{\zeta}\right)\right\}\Psi^{\prime}=0
 \label{PsiEin}
 \end{equation}
where 
\begin{equation}
\tilde{\omega}_{\mu}^{cd}=\omega_{\mu}^{cd}(\tilde{e})+
\frac{1}{4M^{2}_{p}}\eta_{ci}\tilde{e}_{d\mu}\varepsilon^{abcd}
\overline{\Psi}^{\prime}\gamma^{5}\gamma^{i}\Psi^{\prime},
 \label{connEin}
 \end{equation} 
$\omega_{\mu}^{cd}(\tilde{e})$ is the Riemannian part of the connection
and $C^{\prime b}_{ab}$
is the trace of the Ricci rotation coefficients\cite{Gasp} in the new
variables. 
Eq.(\ref{connEin}) coincides with the well-known solution for the spin
connection in the context of the first order formalism approach
to the Einstein-Cartan theory\cite{Gasp} where a Dirac
spinor field is the only source of a non-Riemannian part of the
connection. For details see also \cite{GK4}.

The gravitational equations are of the standard form (\ref{gef})
with 
\begin{equation}
T_{\mu\nu}^{eff}=\frac{\zeta +b_{k}}{\zeta +b_{g}}
\varphi_{,\mu}\varphi_{,\nu}-K\tilde{g}_{\mu\nu}
+\frac{b_{g}M^{4}}{(\zeta +b_{g})^{2}}
e^{-2\alpha\phi/M_{p}}\tilde{g}_{\mu\nu}
+\frac{\zeta}{\zeta +b_{g}}T_{\mu\nu}^{(f,canonical)}
-\frac{mF(\zeta)}{\sqrt{\zeta +b_{g}}}
\overline{\Psi}^{\prime}\Psi^{\prime}\tilde{g}_{\mu\nu},
 \label{Tmn+f}
\end{equation}
where 
\begin{equation}
T_{\mu\nu}^{(f,canonical)}=
\frac{i}{2}[\overline{\Psi}^{\prime}\gamma^{a}e_{a(\mu }^{\prime}
\nabla_{\nu )}\Psi^{\prime}-(\nabla_{(\mu}\overline{\Psi}^{\prime})
\gamma^{a}e_{\nu )a}^{\prime}\Psi^{\prime}]
\label{Tmnfcanon}
\end{equation}
is the canonical energy-momentum tensor for the fermionic field
in the curved space-time\cite{Birrel} and
$\nabla_{\mu}\Psi^{\prime} =\left(\partial_{\mu}+
\frac{1}{2}\tilde{\omega}_{\mu }^{cd}\sigma_{cd}\right)\Psi^{\prime}$
and $\nabla_{\mu}\overline{\Psi}^{\prime}=
\partial_{\mu}\overline{\Psi}^{\prime}-
\frac{1}{2}\tilde{\omega}_{\mu}^{cd}\overline{\Psi}^{\prime}\sigma_{cd}$.

The scalar field $\zeta$ is defined by the constraint (\ref{confermEin}) in
terms of the dilaton and fermion fields as a solution of the fifth degree
algebraic equation that makes finding $\zeta$ in general a very complicate
question. However there are two physically most interesting limiting cases
when   
solving (\ref{confermEin}) is simple enough. The only assumption we will
make about dimensionless parameters of the theory in what follows will be 
that $b_{g}$, $b_{k}$ and $h$ are not too large.

Let us first  analyze the constraint (\ref{confermEin})  
when  the fermionic  density (proportional to
$\overline{\Psi}^{\prime}\Psi^{\prime}$) is very low as compared to the 
contributions of the dilaton potential ($\propto M^{4}e^{-2\alpha\phi/M_{p}}$)
 and kinetic term $K$.
In this limiting case,
 the constraint gives again the expression (\ref{con2})
for $\zeta$. If we assume then the quintessential cosmological solution 
of Sec.IV or the halo dark matter solution of Sec.V where 
$Ke^{2\alpha\phi/M_{p}}=const$, we get a constant value of $\zeta$.
Inserting this value of $\zeta$ into (\ref{PsiEin}) we see that the mass 
of a "test" fermion (that is when we ignore the effect of the fermion 
itself on the quintessential or halo dark matter background) is constant.
Notice that the constant mass of fermions can be different in the
cosmological and in the halo solutions (remind that parameters of the
theory needed for these solutions are also different). Notice, however, 
that if 
$\Delta =0$ (that is $b_{g}=b_{k}$), then $\zeta =b_{g}$ and the mass 
of a "test" fermion is constant for any dilatonic background.

An opposite regime is realized  when the contribution of the fermionic  
density to the constraint (\ref{confermEin})  is very high
as compared to the contributions of the dilaton potential and kinetic term.
In the context of the quintessence model of the present day universe,
this regime corresponds in particular to the normal laboratory conditions
in particle physics. 
 Then according to the
 constraint (\ref{confermEin}), one of the possibilities for this to be 
realized consists in the condition
\begin{equation}
F(\zeta)\equiv \frac{1}{2}\left(\zeta +\frac{2b_{g}h}{\zeta}+3h\right)
\approx 0
\label{F=0}
\end{equation}
from which we find two possible {\it constant} values for $\zeta$
\begin{equation}
\zeta\approx -\frac{3}{2}h\left(1\pm\sqrt{1-\frac{8b_{g}}{9h}}\right)
= const.
\label{zeta12}
\end{equation}
Of course, these solutions have sense only if $b_{g}/h<9/8$.
We see from (\ref{PsiEin}) that two different constants $\zeta$ given by
(\ref{zeta12})
define in general {\it two specific masses} for the fermion. Notice that in the 
special case when $b_{g}=0$, Eq. (\ref{F=0}) is linear in $\zeta$ 
and we obtain therefore only one  effective fermion mass
$m^{eff}_{ferm}\approx \frac{2m}{3\sqrt{3|h|}}$. 

Surprisingly that the same factor $F(\zeta)$ appears in the last terms of 
Eqs. (\ref{phief+ferm}) and (\ref{Tmn+f}). Therefore, in the same regime 
of fermion dominance, the last  terms of 
Eqs. (\ref{phief+ferm}) and (\ref{Tmn+f}) {\it automatically} vanish.
In Eq. (\ref{phief+ferm}), this means that the fermion density
 $\overline{\Psi}^{\prime}\Psi^{\prime}$ is not a source for the dilaton 
and thus the long-range force disappears automatically. Notice that 
there is no need to require no interactions of the dilaton with barionic
matter at all to have agreement with observations but  it is rather enough
that these interactions vanish in the appropriate regime where barionic
matter dominates over other matter fields. In Eq. (\ref{Tmn+f}), the condition
(\ref{F=0})
means that in the  region where the fermionic matter dominates,
 the fermion energy-momentum
tensor, up to a constant, becomes equal to the canonical 
energy-momentum tensor of a fermion field in GR  
\footnote{The decoupling of the dilaton in the CEF in the case of high 
fermion density was discussed also in a simpler scale invariant model
(with $b_{g}=b_{k}=0$ and explicit exponential potentials) in Ref. 
\cite{G1}.}.

The separate possibility relevant to the very high fermionic  density 
(again, as compared to the contributions of the dilaton potential and 
kinetic term) is the case when 
\begin{equation}
\zeta +b_{g}\approx 0
\label{zeta=0}
\end{equation}
is a solution. Then $F(\zeta)\approx F(-b_{g})=h-b_{g}$, \quad $\zeta
-b_{g}\approx -2b_{g}$ and it follows from the constraint
(\ref{confermEin})
\begin{equation}
\frac{1}{\sqrt{\zeta +b_{g}}}\approx 
\frac{m}{4M^{4}}\left(\frac{h}{b_{g}}-1\right)
\overline{\Psi}^{\prime}\Psi^{\prime}e^{2\alpha\phi /M_{p}}.
\label{srtzeta}
\end{equation}

In this case, Eq. (\ref{PsiEin}) describes fermion with an effective
quartic self-interaction like in NJL model\cite{NJL}. The coupling
constant of this self-interaction depends on the dilaton $\phi$.
For example, the condition (\ref{zeta=0}) is realized as 
$\phi\rightarrow\infty$ that corresp[onds to the late universe in 
the quintessence scenario.

Here, as opposed to the solutions (\ref{zeta12}), we get quartic
interaction instead of mass generation. We expect however that after 
$\overline{\Psi}^{\prime}\Psi^{\prime}$ develops an expectation value,
mass
generation will be possible as in NJL model \cite{NJL} (for recent
progress in this subject see e. g. Ref. \cite{Cv}). It is interesting to
note that appearance of the quartic self-interaction here is related to the
SSB of the scale invariance. In fact, Eq. (\ref{srtzeta})  tells us that
without SSB of
scale invariance such quartic interaction is not defined.

Concluding this analysis of equations  when  the fermionic density is of 
the order
typical for the normal particle physics (which in the laboratory 
conditions is always much higher than the dilaton density )
we see
that starting from a single
fermionic field we obtain (if $b_{g}\neq 0$) exactly three different types
of spin $1/2$
particles in CEF. This appears to be a new approach to the family problem
in particle physics. This is why we will refer to the described effect 
as the "family birth effect". 

All what has been done here concerning fermions is in the context of a toy
model without Higgs fields, gauge bosons and the associated
$SU(2)\times U(1)\times SU(3)$ gauge symmetry of the standard model. As we have seen
in other models (see \cite{GK4}, the second reference of \cite{G1} and
\cite{K}), it is possible to incorporate the two measure ideas  with the 
gauge symmetry and Higgs mechanism. Now the differences consist of:
i) the presence of global scale symmetry, ii) the most general TMT
structure for gravitation and dilaton sector but including only kinetic terms.
The complete discussion of the standard model in the context of such TMT
structure will be presented in a separate publication\cite{prep}.
Here we want only to explain shortly the main ideas that provides us the
possibility to implement this program.

It is important that in a simple way gauge fields can be incorporated so
that they will not appear in fundamental constraint\footnote{This may be
done by making the gauge field kinetic terms coupled to $\sqrt{-g}$ and
the Higgs field kinetic term coupled to measure $\Phi$. Both of these 
things are dictated also by local scale invariance of that part of the
action.}     
in contrast to the fermions (see for comparison Eq. (\ref{confermEin})).
As it is easy to see, the different constant values of $\zeta$
corresponding
to the solutions of the constraint do
not change the expectation value of the Higgs field. We can also work
without significant changes in the discussion of the fermionic sector if
instead of explicit mass-like terms we will work with similar terms where
the coupling constants with the dimensionality of the mass   are replaced
by
gauge invariant Yukawa couplings to the Higgs field. Once again we find 
three fermion families, as was done above in the toy model. Generating
mass of two of them is automatic as in the previous discussion. For the
third we need again some quantum effect that gives rise to expectation
value of the gauge invariant Yukawa coupling terms. Since the object that
gets expectation  value is gauge invariant, we don't expect further
breaking of gauge symmetry (as opposed to usual analysis on top quark
condensates\cite{Cv})\footnote{ Notice that in our case there is an
explicit Higgs field as opposed to the top quark
condensate models, and in spite of this we need the condensates of the
Yukawa coupling terms so as to get a normal mass term for the third
family.}

\section{Discussion and conclusions}

In this paper the possibility of a spontaneously generating exponential
potential for the dilaton field in the context of TMT with spontaneously
broken global scale symmetry was studied. The symmetry transformations
formulated
in terms of the original variables (\ref{st}) (or (\ref{ctferm}) in the
presence of fermions) include the global scale transformations of the
metric,
of the scalar fields $\varphi_{a}$ related to the measure $\Phi$ (and
of the fermion fields) and in addition the dilaton field $\phi$ undergoes
a global shift. In the CEF  (see Eqs. (\ref{ct}) or 
(\ref{ctferm}) where
the theory is
formulated in the Riemannian (or Einstein-Cartan) space-time), all
dynamical variables are invariant under the transformations (\ref{st}) (or
(\ref{ctferm})) except for the dilaton field which still gets shifted by a
constant. Thus, SSB of the scale symmetry that appears firstly in
(\ref{varphi}) when solving Eq. (\ref{7}), is reduced, in the CEF,
 to SSB of the shift symmetry   $\phi\rightarrow\phi
+const$.   

The original action does not includes potentials but in the
CEF, the exponential potential
appears as a result of SSB of the scale symmetry. In the generic case
$\Delta
=b_{g}-b_{k}\neq
0$, the process of SSB also produces terms with 
higher powers in derivatives of the dilaton field.

Cosmological scaling solutions of the theory were studied.
The flatness of the potential $V_{eff}$ which is associated here with the
exponential form, is protected by the scale symmetry.
 Quintessence
solutions (corresponding to accelerating universe) were found possible
 for
a range of parameters if the integration constant $sM^{4}$ in Eq.
(\ref{varphi}) 
is chosen to be positive. 

Also in the same model, but for a  negative
integration constant $sM^{4}$ and for a different range of the parameter
$\alpha$ it is found that halo-like solutions exist. They give rise to a
constant velocity for test particles moving at large distances
in circular orbits.

Finally, the behavior of fermions in such type of models was investigated.
Scale invariant fermion mass-like terms can be introduced in two different
ways since they can appear coupled to each of the two different measures
of the theory. Although an exponential of the dilaton field $\phi$
couples to the fermion in both of these terms, it is found that when the
fermions are treated as a test particles in the scaling background, their
masses in the CEF are constant.

Even more surprising is the behavior of the fermions in the limit of high
fermion density as compared to the dilaton density. This approximation is
regarded as more realistic if we are interested in the regular particle
physics behavior of these fermions under normal laboratory conditions.
It is found then that in the CEF, a given fermion can
behave in three different ways
according to the three different solutions of the fundamental constraint
(\ref{confermEin}). Two of the solutions correspond to fermions with
constant masses and the other - to a NJL model\cite{NJL}, which is known
can generate mass on the quantum level. From one fermion three are
obtained for free. This suggests a new approach to the "family problem" in
particle physics.

In addition to this, for the two mentioned above solutions (\ref{zeta12}) 
corresponding to constant fermion masses, the fermion-dilaton coupling 
in the CEF (proportional to $F(\zeta)$,
Eq.(\ref{F})) disappears automatically. If one of these types of fermions
is associated to the first family (regular matter, i.e., $u$ and $d$
quarks, $e^{-}$ and $\nu_{e}$), we obtain that normal matter decouples
from the dilaton. 

The analysis of the constraint (\ref{confermEin}) in the case where the 
fermionic density is of the same order as the dilaton energy density will 
provide in general five solutions for $\zeta$. It could be that those
"low energy families" may be a good candidate for dark matter.

\section{Acknowledgments}

We are grateful to
J. Alfaro, S. de Alwis, M. Banados, J. Bekenstein,  Z. Bern, R. Brustein,  
K. Bronnikov, V. Burdyuzha,
L. Cabral, S. del Campo, C. Castro, 
A. Davidson, M. Eides,  P. Gaete, H. Goldberg, M. Giovannini, A. Guth, 
F. Hehl,
V. Ivashchuk, A. Linde,
L. Horwitz, M. Loewe, A. Mayo, Y. Ne'eman, Y. Jack Ng,
M. Pavsic, M.B.Paranjape, L. Parker,
J. Portnoy, V.A. Rubakov,
E. Spallucci, Y. Verbin, A. Vilenkin, M. Visser, K. Wali, C. Wetterich, 
P. Wesson
and J.Zanelli for discussions on different aspects of this paper.


\bigskip
\begin{thebibliography}{99}

\bibitem{P} 
See, for example, N. Bahcall, J.P. Ostriker, S.J. Perlmutter and
P.J. Steinhardt, Science {\bf 284}, 1481 (1999) and references therein.

\bibitem{Wett1988NP668}
C. Wetterich, Nucl. Phys.  {\bf B302}, 668 
(1988); B. Ratra and P.J.E. Peebles , Phys. Rev. {\bf D37}, 3406 (1988);
P.J.E. Peebles and B. Ratra, Astrophys. J. {\bf 325}, L17
(1988); R. Caldwell, R. Dave and P. Steinhardt, Phys. Rev. Lett. {\bf 80},
1582 (1998); N. Weiss, Phys. Lett. {\bf B197}, 42 (1987); 
Y. Fujii and T. Nishioka, Phys. Rev. {\bf D42}, 361 (1990);
M.S. Turner and M. White, Phys. Rev. {\bf D56}, R4439 (1997);
E. Copeland, A. Liddle and D. Wands, Phys. Rev. {D57}, 4686 (1998);
C.T. Hill, D.N. Schramm and J.N. Fry, Comments Nucl. Part. Phys. {\bf 19},
25 (1989); J. Frieman, C. Hill and R. Watkins, Phys. Rev. {\bf D46}, 1226
(1992); J. Frieman, C. Hill, A. Stebbins and I. Waga, Phys. Rev. Lett.
{\bf 75}, 2077 (1995); C. Wetterich, Nucl. Phys. {\bf B302}, 645 (1988); 
Astron. Astrophys. {\bf 301}, 321 (1995);
P. Ferreira and M.Joyce, Phys. Rev. Lett. {\bf 79},
4740 (1997);
I. Zlatev, L. Wang and P. Steinhardt, Phys. Rev. Lett.
{\bf 82}, 896 (1999);
P. Steinhardt, L. Wang  and I. Zlatev, Phys. Rev. {\bf D59}, 123504 (1999). 

\bibitem{FJ}
P. Ferreira and M.Joyce, Phys. Rev. {\bf D58}, 023503 (1998).


\bibitem{CCP}
I. Novikov, {\it Evolution of the Universe,}
Cambridge University Press, 1983.
S.Weinberg, Rev. Mod. Phys. {\bf 61}, 1 (1989) and 
{\it The Cosmological Constant Problem}, astro-ph/0004075;
 Y.J. Ng, Int. J. Mod.
Phys. {\bf D1}, 145, (1992); 
S.M. Carroll, W.H. Press and E.L. Turner, Ann. Rev. Astron. and Astrophys.
{\bf 30},499 (1992); V. Sahni and A. Starobinsky, "The Case for a Positive
Cosmological $\Lambda$-term", astro-ph/9904398.
  S.M. Carroll, {\it The 
cosmological constant}, astro-ph/0004075; P. Binetruy, {\it Cosmological
constant vs. quintessence}, hep-ph/0005037; J. Carriga and A. Vilenkin,
{\it Solutions to the cosmological constant problems}, hep-th/0011262.


\bibitem {KLyth}
C. Kolda and D. Lyth, Phys. Lett {B458}, 197 (1999).


\bibitem{Carroll}
S.M. Carroll, Phys. Rev. Lett. {\bf 81}, 3067 (1998).

\bibitem{Dolgov}
A.D. Dolgov, Phys. Reports {\bf 320}, 1 (1999).

\bibitem{GK1}
E.I. Guendelman and A.B. Kaganovich, Phys. Rev. {\bf D53}, 7020
(1996); Mod. Phys. Lett. {\bf A12}, 2421
(1997); Phys. Rev. {\bf D55}, 5970 (1997). 

\bibitem{GK2}
E.I. Guendelman and A.B. Kaganovich, Phys. Rev.{\bf D56}, 3548 (1997).
 

\bibitem{GK3}
E.I. Guendelman and A.B. Kaganovich, Phys. Rev.{\bf D57}, 7200 (1998);
Mod. Phys. Lett. {\bf A13}, 1583 (1998) 

\bibitem{GK4}
E.I. Guendelman and A.B. Kaganovich, Phys. Rev. {\bf D60}, 065004 (1999).


\bibitem{G}
E.I. Guendelman, Mod. Phys. Lett. {\bf A14}, 1043 (1999); 
 Class. Quant. Grav. {\bf 17}, 361 (2000);
gr-qc/0004011. 

\bibitem{G1}
E.I. Guendelman, Mod. Phys. Lett. {\bf A14}, 1397 (1999); gr-qc/9901067;
hep-th/0106085.

\bibitem{G2}
E.I. Guendelman, hep-th/0011049, to appear in Foundations of Physics.

\bibitem{K}
A.B. Kaganovich, Phys. Rev.{\bf D63}, 025022 (2001).

\bibitem{PV}
P.J.E. Peebles and A. Vilenkin, Phys. Rev. {\bf D59}, 063505 (1999).

\bibitem{exppot1}
C. Wetterrich, Nucl. Phys. {\bf B252}, 309 (1985); J. Halliwell, ibid. 
{\bf B266}, 228 (1986); Q. Shafi and C. Wetterrich, Phys. Lett. 
{\bf 152B}, 51 (1985); Q. Shafi and C. Wetterrich, Nucl. Phys. {\bf B289},
787 (1987).

\bibitem{exppot2}
E. Cremer et al., Phys. Lett. {\bf 133B}, 61 (1983); J. Ellis et al.,
ibid. {\bf 134B}, 429 (1984); H. Nishino and E. Sezgin, ibid {\bf 144B},
187 (1984); E. Witten, ibid. {\bf 155B}, 151 (1985); M. Dine et al.,
ibid. {\bf 156B}, 55 (1985); J. Halliwell, ibid. {\bf 185B}, 341 (1987).
J. Yokoyama and M. Maeda, ibid. {\bf 207B}, 31 (1988). 

\bibitem{exppot3}
J. Barrow and S. Cotsakis, Phys. Lett., {\bf 214B}, 515 (1988); 
S. Cotsakis, Phys. Rev. {\bf D47}, 1437 (1993).

\bibitem{WP}
D. Gross and E. Witten, Nucl. Phys. {\bf B277}, 1 (1986); 
J. Polchinski, {\it Superstrings}, Vol.II (Cambridge U. Press, 
Cambridge, 1998), Ch. 12.

\bibitem{BV}
M. Barriola and A. Vilenkin, Phys. Rev. Lett. {\bf D63}, 341 (1989).

\bibitem{Nuc}
U. Nucamendi, M. Salgado and D. Sudarsky, Phys. Rev. Lett. {\bf D84}, 341
(2000); gr-qc/0011049.

\bibitem{Guz}
T. Matos, F. S. Guzman and D. Nunez
Phys. Rev. {\bf D62}, 061301 (2000).


\bibitem{Gasp}
V. de Sabbata and M. Gasperini, {\it Introduction to Gravity} (World
Scientific, Singapore, 1985).


\bibitem{Birrel}
N.D. Birrell and P.C.W. Davies, {\it Quantum fields in curved space}
(Cambridge University Press, 1982). 

\bibitem{NJL}
Y. Nambu and G. Jona-Lasinio, Phys. Rev. {\bf 122}, 345.

\bibitem{Cv}
G. Cvetic, Rev. Mod. Phys. {\bf 71}, 513 (1999).

\bibitem{prep}
E.I. Guendelman and A.B. Kaganovich, in preparation.

\end{thebibliography}
\end{document}